\documentclass{iaus}
\usepackage{graphicx}

\pdfoutput=1

\title[Coexistence of Classical Bulges and Pseudobulges] %% give here short title %%
{The Coexistence of Classical Bulges and Disky Pseudobulges in Early-Type 
Disk Galaxies}

\author[Erwin]   %% give here short author list %%
{Peter Erwin$^{1,2}$}%
\affiliation{$^1$Max-Planck-Insitut f\"{u}r extraterrestrische Physik, 
Giessenbachstrasse, 85748 Garching, Germany \break email: erwin@mpe.mpg.de\\[\affilskip]
$^2$Universit\"{a}ts-Sternwarte M\"{u}nchen, Scheinerstrasse 1,
81679 M\"{u}nchen, Germany}

\pubyear{2007}
\volume{245}  %% insert here IAU Symposium No.
\pagerange{000--000}
\date{?? and in revised form ??}
\jname{Formation and Evolution of Galaxy Bulges}
\editors{M. Bureau, A. Athanassoula, \& B. Barbuy, eds.}
\begin{document}

\maketitle

\begin{abstract}
Close examination of ``pseudobulges'' in several early-type disk galaxies
indicates that they are actually composite structures consisting of both a
flattened, kinematically cool disklike structure (``disky pseudobulge'') and a
rounder, kinematically hot spheroidal structure (``classical bulge'').  This
indicates that pseudobulges, thought to form from internal secular evolution,
and classical bulges, thought to form from rapid mergers, are not exclusive
phenomena: some galaxies can have both.
\end{abstract}

\firstsection % if your document starts with a section,
\section{Introduction}

There has been considerable debate recently on the issue of ``pseudobulges'':
cases where the central ``bulge'' of a disk galaxy is apparently \textit{not}
a small elliptical galaxy embedded within the disk (the ``classical'' model of
a bulge) but instead something more disklike and -- it is thought -- the
result of internal, secular evolution processes instead of violent early
mergers (\cite[see the review by Kormendy \& Kennicutt 2004]{kk04}).  The
tendency has been to contrast pseudobulges and classical bulges as in some
sense exclusive categories: galaxies have either a classical bulge (e.g.,
early-type disks) or a pseudobulge (e.g., late-type disks), but not both at
once (\cite[but see Athanassoula 2005]{athan05}).  Here, I present photometric
and kinematic evidence suggesting that some galaxies \textit{can} have both at
once.

\section{Definitions (Or, What \textit{I} Mean by ``Bulge'' and
``Pseudobulge'')}

The term ``pseudobulge'' has become overly general and confused, with
different authors using the term to describe different stellar structures, and
using different criteria to identify them.  In this study, I recognize and am
concerned with three entities:

\begin{itemize}
  \item \textbf{Photometric Bulge:} This is the region of a galaxy defined by
  a bulge-disk decomposition, where excess light in the inner regions of a
  galaxy dominates over the light of the outer exponential disk.
  
  \item \textbf{Classical Bulge:} This is the traditional idea of a ``bulge'':
  a spheroidal or weakly triaxial collection of stars dominated by random
  motions (velocity dispersion).  It can be recognized by being significantly
  rounder (in projection) than the disk, and by being kinematically hot.
  
  \item \textbf{(Disky) Pseudobulge:} This is an inner disk of stars, with a
  geometry (flattening) similar to that of the outer disk.  A ``kinematic''
  pseudobulge is one with evidence for rotationally dominated stellar
  kinematics (i.e., more like disk kinematics).  Secondary characteristics can
  include nuclear bars, spirals, rings, etc., but these are not required.
\end{itemize}

Note that I do not consider ``boxy/peanut-shaped bulges,'' which are well
understood as the vertically thickened inner part of bars.  I also do not
touch on issues of stellar colors, star formation, or dust; S0 galaxies have
been shown to harbor disky pseudobulges even though they are generally free of
dust and recent star formation (\cite[Erwin, Vega Beltr\'an, Graham, \etal\
2003]{erwin03}).  Finally, I do not assume that particular surface-brightness
profiles (as indicated by, e.g., S\'ersic indices) necessarily belong to one
category or another, although in practice disky pseudobulges often have
quasi-exponential profiles.

\section{Methodology}

The basic approach is to first identify the photometric bulge of a galaxy via
a bulge-disk decomposition.  The morphology, structure, and geometry of the
photometric bulge is then examined, to see if it is predominately disklike or
spheroidal.  Finally, the stellar kinematics of the photometric bulge are used
to determine if the stellar motions are dominated by velocity dispersion (as
for a classical bulge) or by rotation.

Kinematically, photometric bulges can be judged by placing them on the
well-known $V_{\rm max}/\sigma$ diagram.  Empirically, almost all ellipticals
-- and a number of unambiguously classical bulges in edge-on spirals -- fall
on or below the line traced by an idealized isotropic oblate rotator model
(IOR; \cite[Binney 1978]{binney78}).  Following \cite{k82}, I identify
photometric bulges which fall \textit{above} the IOR curve as kinematic
pseudobulges: they are dominated by rotation to a degree not seen in
ellipticals and bona fide classical bulges.

\begin{figure}
\includegraphics[scale=0.62]{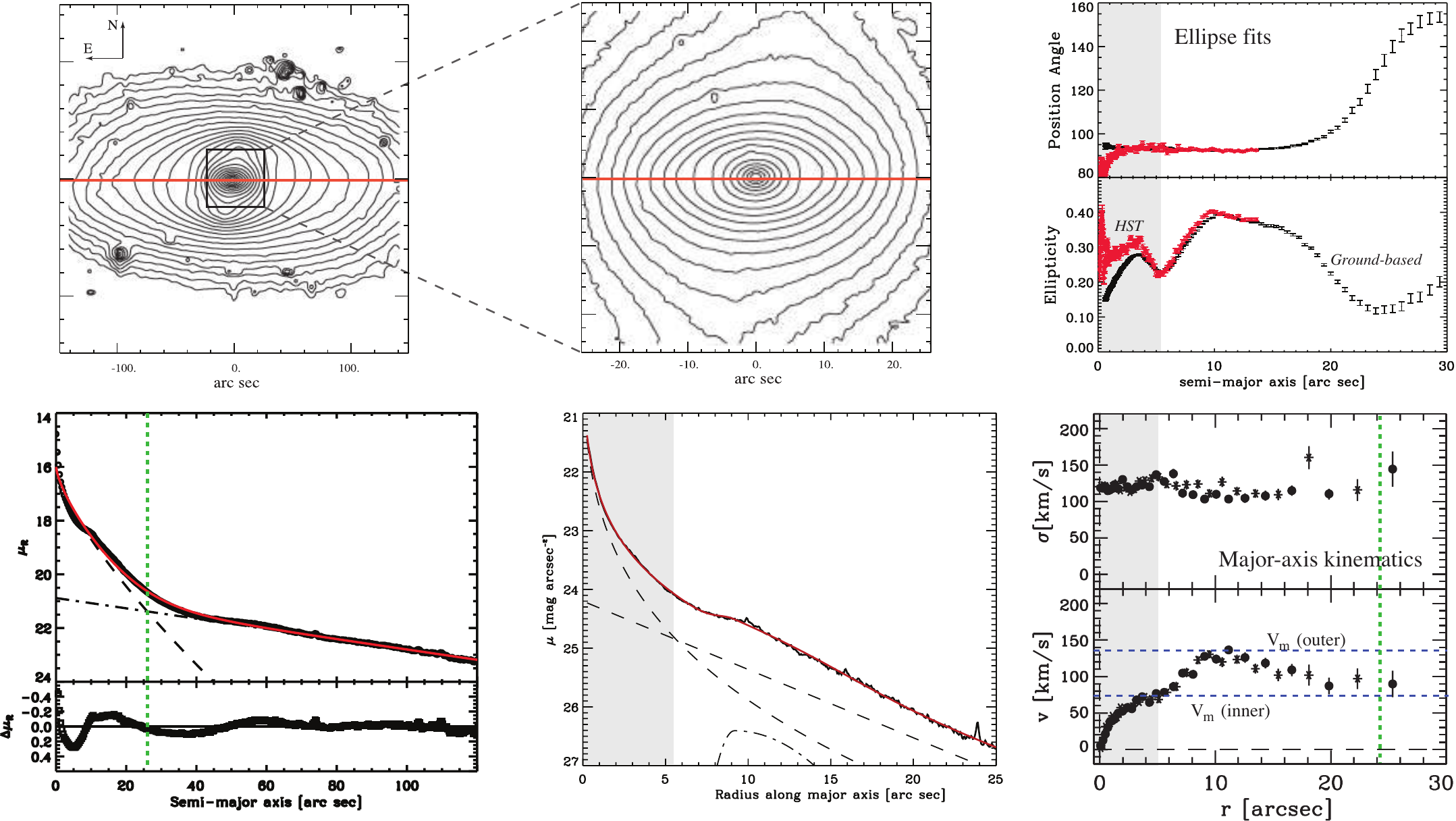}
	\caption{Upper left: $R$-band isophotes of NGC 4371, with major-axis cut
	line.  Lower left: Major-axis surface-brightness profile, with S\'ersic +
	exponential fit.  The S\'ersic-dominated region (the ``photometric
	bulge'') is $r < 30^{\prime\prime}$; the vertical dashed green line marks
	the boundary.  Upper middle: zoom of the photometric bulge region, showing
	elliptical isophotes and rounder isophotes inside.  Lower middle:
	Surface-brightness profile of the photometric bulge region, fit with
	S\'ersic + exponential + ring; an ``inner photometric bulge'' dominates
	the light at $ r < 6^{\prime\prime}$ (light gray shading).  Upper right:
	Ellipse fits to ground-based and HST isophotes.  Lower right: Major-axis
	stellar kinematics.  The stellar velocity has a clear maximum in the
	photometric bulge region ($r \sim 10^{\prime\prime}$), and a plateau in
	the \textit{inner} photometric bulge ($r \sim
	4^{\prime\prime}$).}\label{fig:n4371}
\end{figure}

All of this follows in the path pioneered by \cite{k82}.  What is new is the 
idea of a recursive approach once a disky pseudobulge has been identified.  
By using high-resolution imaging (e.g., from HST) and spectroscopy, we can 
focus on the inner region of the pseudobulge and -- in some cases, at least 
-- identify a separate ``inner photometric bulge'' which has rounder 
isophotes and hotter kinematics.

% \begin{figure}
% \includegraphics[scale=0.55]{erwin_fig2}
%    \caption{NGC 4371: Inside the photometric bulge.  Upper left:
%     Surface-brightness profile of the photometric bulge region, fit with
%     S\'ersic + exponential + ring.  Lower left: Ellipse fits to ground-based
%     and HST isophotes.  Upper right: zoom of the photometric bulge region,
%     showing rounder isophotes in the center.  Lower right: Kinematic profile
%     from Figure~1, now focused on the inner region.}\label{fig:n4371inner}
% \end{figure}
% 

\section{Example: NGC 4371}

% I proceed in two stages.  The first stage is the identification of the
% photometric bulge region as being at least partly disklike (\cite[e.g., Erwin
% \etal\ 2003]{erwin03}).  To being, I isolate the photometric bulge region by
% doing a bulge-disk decomposition; the photometric bulge is defined the region
% where the ``bulge'' (S\'ersic) component is brighter than the exponential disk
% component.  For NGC 4371, this $r < 25^{\prime\prime}$ (Figure~1).  Next, I
% examine the

NGC~4371 is a barred S0 galaxy in the Virgo Cluster, notable for a bright
stellar nuclear ring inside its bar (\cite[Erwin \& Sparke 1999]{erwin99}).
In the left-hand panels of Figure~\ref{fig:n4371}, I show a major-axis
surface-brightness profile, along with a simple bulge-disk decomposition.
This decomposition defines the photometric bulge as the region at $r <
30^{\prime\prime}$.  Inspection of the isophotes (upper middle and upper right
panels of Figure~\ref{fig:n4371}) shows that this region is precisely where
the stellar nuclear ring is found.  The ellipticity reaches a peak of
$\epsilon = 1 - b/a = 0.40$ at $a \sim 10^{\prime\prime}$; this is close to
the ellipticity of the \textit{outer} disk (0.45).

\begin{figure}
\includegraphics[scale=0.72]{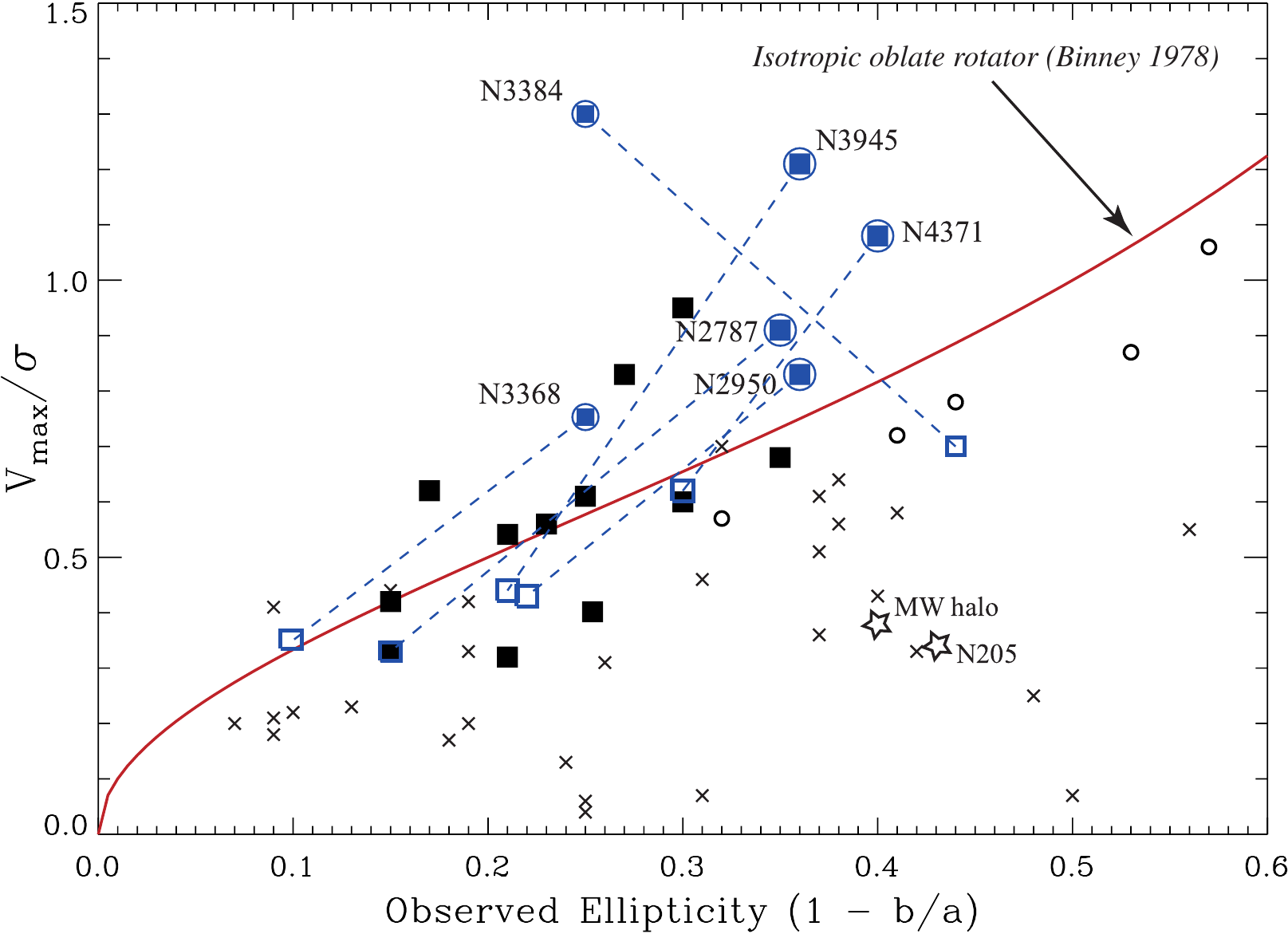}
   \caption{$V_{\rm max}/\sigma$ diagram.  Crosses are elliptical galaxies
   (\cite[Davies \etal\ 1983]{davies83}), open circles are large, edge-on
   bulges (\cite[Kormendy \& Illingworth 1982; Jarvis \& Freeman
   1985]{ki82,jarvis85}), and stars are dE galaxy NGC 205 (\cite[Geha et al.\
   2006]{geha06}) and the Milky Way halo (\cite[Ibata \etal\ 2007 and
   references therein]{ibata07}).  Filled black squares are photometric bulges
   of barred S0 galaxies (Erwin \etal\ 2008); circled/filled blue squares are
   disky pseudobulges with inner classical bulges (open blue squares), with
   dashed lines connecting classical bulges and disky pseudobulges of the same
   galaxy.}\label{fig:vsigma}
\end{figure}

The stellar kinematics (Figure~\ref{fig:n4371}, lower right) show a clear
velocity maximum at roughly the same radius as the nuclear ring.  The ratio of
this velocity to the mean velocity dispersion is $\approx 1.1$, well above the
IOR curve given its apparent ellipticity (predicted $V_{\rm max}/\sigma =
0.82$; Figure~\ref{fig:vsigma}).  Since the photometric bulge is both highly
flattened (like a disk) and dominated by rotation (kinematically similar to a
disk), it qualifies as a pseudobulge.  (This was originally pointed out by
\cite[Kormendy 1982]{k82}.)

However, close inspection of the pseudobulge using HST images shows that at
small radii ($r < 6^{\prime\prime}$), the isophotes becomes distinctly
\textit{rounder}, with $\epsilon \approx 0.3$.  (The minimum in ellipticity
just outside this zone is consistent with the isophotal effects of an
elliptical ring superimposed on a rounder bulge; see \cite[Erwin \etal\
2001]{erwin01}).  A bulge/disk decomposition of the photometric bulge profile
shows that it is well fit by an exponential + an inner S\'ersic component
(along with an asymmetric Gaussian to account for the ring; lower middle panel
of Figure~\ref{fig:n4371}).  The resulting ``inner'' photometric bulge turns
out to be precisely that region ($r < 6^{\prime\prime}$) where the isophotes
become rounder.  So it appears that while the outer part of the pseudobulge is
highly flattened and nearly exponential in profile, the interior harbors an
additional, rounder component.  The kinematics in \textit{this} region (lower
right panel, Figure~\ref{fig:n4371}) have $V_{\rm max}/\sigma$ almost
identical to the $V_{\rm max}/\sigma$ predicted for $\epsilon = 0.3$.  The
innermost component thus appears to be a classical bulge: rounder than the
pseudobulge (and the outer disk) and \textit{not} dominated by rotation.

\section{Discussion}

Preliminary analyses have been done for five more galaxies (four S0 and one
Sab) which show similar composite structures and kinematics; these are
indicated by blue symbols in Figure~\ref{fig:vsigma}.  Note that since these
galaxies have bars, they most likely also have the box/peanut structures of
bars; these galaxies probably have all three of the distinct components
suggested by \cite{athan05}.  Comparison with an ongoing study of pseudobulges
in barred S0 galaxies (\cite[Erwin, Aguerri, Beckman, \etal\ 2008]{erwin08})
suggests that at least one quarter of S0 pseudobulges may be composite
systems.

Two of these composite systems (NGC~2787 and NGC~3384) are known supermassive
black hole hosts, which suggests a possible resolution to an issue first
raised by \cite{kg01}.  Black hole masses correlate well with properties of
host elliptical galaxies, and also with properties of host (photometric)
bulges in disk galaxies, which suggests common evolutionary mechanisms.
However, black holes are apparently also found in galaxies with pseudobulges,
which are thought to have formation mechanisms different from those of
ellipticals and classical bulges.  If pseudobulge galaxies with black holes
\textit{also} have classical bulges, then the idea of a common evolutionary
mechanism driving black hole and (classical) bulge growth may still be
tenable.

%\begin{discussion}

%\end{discussion}

\end{document}